\documentclass[a4paper]{report}
\usepackage[utf8]{inputenc}
\usepackage[T1]{fontenc}
\usepackage{RJournal}
\usepackage{amsmath,amssymb,array}
\usepackage{booktabs}

\begin{document}

\sectionhead{Contributed research article}
\volume{XX}
\volnumber{YY}
\year{20ZZ}
\month{AAAA}

\begin{article}
  \title{R package \pkg{imputeTestbench} to compare imputations methods for univariate time series}
\author{by Neeraj Bokde, Kishore Kulat, Marcus W Beck, Gualberto Asencio-Cort\'{e}s}

\maketitle

\abstract{
This paper describes the R package \pkg{imputeTestbench} that provides a testbench for comparing imputation methods for missing data in univariate time series. The \pkg{imputeTestbench} package can be used to simulate the amount and type of missing data in a complete dataset and compare filled data using different imputation methods.  The user has the option to simulate missing data by removing observations completely at random or in blocks of different sizes.  Several default imputation methods are included with the package, including historical means, linear interpolation, and last observation carried forward.  The testbench is not limited to the default functions and users can add or remove additional methods using a simple two-step process. The testbench compares the actual missing and imputed data for each method with different error metrics, including \textit{RMSE}, \textit{MAE}, and \textit{MAPE}. Alternative error metrics can also be supplied by the user. The simplicity of use and significant reduction in time to compare imputation methods for missing data in univariate time series is a significant advantage of the package. This paper provides an overview of the core functions, including a demonstration with examples.
}

\section{Introduction}

The CRAN repository includes many packages for imputing missing values. These packages have been used in a wide range of applications including biomedical research, civil and urban planning, design of medical care systems, and ecological research. Several packages are well recognized and easy to use, including \CRANpkg{MICE}, \CRANpkg{mi}, \CRANpkg{Amelia}, and \CRANpkg{missForest}.  The \pkg{MICE} \citep{buuren2011mice} (Multivariate imputation via Chained equation) package provides several imputation methods for time series with data that are Missing at Random (\textit{MAR}).  Imputation methods used by \pkg{MICE} include Predictive mean matching (\textit{PMM}), logistic regression, Bayesian polytomous regression, and proportional odds model. The \pkg{mi} (Multiple imputation with diagnostic) \citep{su2011multiple} package also uses the predictive mean matching technique, whereas bootstrapping and the EMB algorithm can be used for \textit{MAR} data in the \pkg{Amelia} package \citep{honaker2011amelia}. An alternative approach is provided by the \pkg{missForest} package \citep{stekhoven2012missforest} that uses a Random Forest algorithm for imputation. Several other methods and packages are provided  by CRAN as discussed in the \href{https://cran.r-project.org/web/views/OfficialStatistics.html}{CRAN Task View: Official Statistics \& Survey Methodology}. Given the amount and types of available methods, the \pkg{imputeTestbench} R Package is proposed as a testbench for efficient comparisons to inform the use of imputation methods.

Studies that have compared imputation methods have used similar approaches to evaluate the superiority of one method over another. For example, \citet{zhu2011missing} proposed a kernel-based iterative estimation method for missing data and compared this method to a non-parametric iterative signal kernel method, non-parametric iterative signal with an RBF kernel, the traditional kernel non-parametric missing value method, and other conventional frequency estimators. The methods were compared by simulating different amounts of missing data, predicting the missing values with each method, and then comparing the predictions to the removed data using the standard root-mean squared error (\textit{RMSE}).  Table \ref{tab:label1} reproduces the results, where the rows show \textit{RMSE} for each imputation method at 10\% and 80\% missing data.

\begin{table}
\centering
\begin{tabular}{lcccc}
\toprule

&\multicolumn{2}{c}{\bfseries 10\%}&\multicolumn{2}{c}{\bfseries 80\%}\\
Method names& \textit{T}&\textit{V}&\textit{T}&\textit{V}\\

\midrule
Mixing&8&0.085&20&1.53\\

Poly&10&0.103&25&2.11\\

RBF&11&0.107&29&2.86\\

Normal&14&0.121&30&3.01\\

FE&13&0.117&29&2.59\\

\bottomrule
\end{tabular}
\caption{Comparison of imputation methods by varying the amount of missing data (10\% and 80\%) and number of iterations. Reproduced from \citet{zhu2011missing}. \textit{T} is the number of iterations for each imputation method and \textit{V} is the mean \textit{RMSE} of the imputed values.}
\label{tab:label1}
\end{table}

\begin{figure}[htbp]
  \centering
  \includegraphics{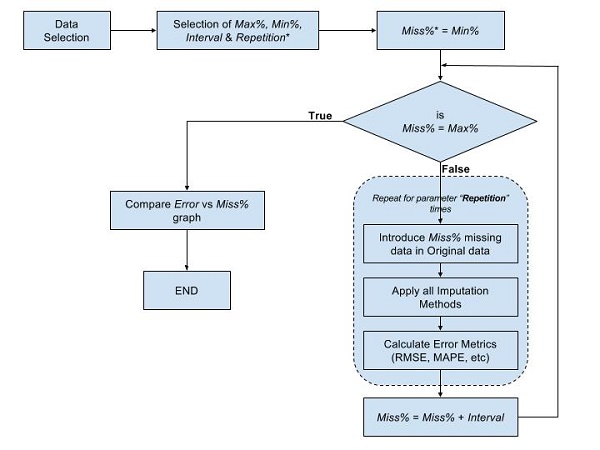}
  \caption{Workflow diagram for comparing imputation methods. \textit{Min}\% and  \textit{Max}\% are minimum and maximum percent of missing values in the dataset. Details are described below.}
  \label{figure:bd}
\end{figure}

Similarly, \citet{tak7444178} proposed an imputation method based on a modified k-nearest neighbour approach that accounted for the effects of spatial and temporal correlation between observations. Missing observations were simulated by removing values from $0.1$\% to $50$\% of the complete data, and then imputed with the proposed method, the nearest history (NH) method, bootstrapping based expectation maximization (B-EM), and the maximum likelihood estimation (MLE) method. The imputation methods were compared using \textit{RMSE}, mean absolute percent error (\textit{MAPE}), and percent change in variance (\textit{PCV}). Additional comparisons in \citet{oh2011biological,jornsten2007meta,li2015hybrid,nguyen2013diagnosing,sim2015missing,li2004towards,ran2015traffic} have used a similar workflow to compare the performance of imputation methods for missing data. This general procedure is summarized in Figure \ref{figure:bd}. The \pkg{imputeTestbench} package formalizes this approach by providing several functions that can greatly simplify the comparison of imputation methods. 

\section{Overview of \pkg{imputeTestbench}}

This section introduces the \CRANpkg{imputeTestbench} R package. The package \pkg{imputeTestbench} \citep{imputeTB} can be used to evaluate imputation methods for univariate time series by simulating missing data and comparing the predictions to the actual.  Previous analyses have manually evaluated the performance of different methods by noting the errors at different percentages of missing values for several repetitions with different error metrics. This approach can be time consuming and prone to errors. Moreover, the relative accuracies of different performance evaluations is a concern given the unavailability of a common platform for comparison. 

The \pkg{imputeTestbench} package is introduced to address the above issues.  The package imports \CRANpkg{dplyr} \citep{dplyr}, \CRANpkg{forecast} \citep{forecast}, \CRANpkg{ggplot2} \citep{ggplot2}, \CRANpkg{imputeTS} \citep{imputeTS}, \CRANpkg{reshape2} \citep{reshape2}, \CRANpkg{tidyr} \citep{tidyr}, and \CRANpkg{zoo} \citep{zoo}. Five relevant functions are included in \pkg{imputeTestbench}.  The primary function is \code{impute\_errors()} which is used to evaluate different imputation methods for missing data that are randomly generated from a complete dataset.  The method for generating missing data for imputation in the test or user-supplied dataset is of particular importance and different methods are provided by the \code{sample\_dat()} function.  The evaluation methods for the imputed data are in the \code{error\_functions()} function.  The remaining two functions, \code{plot\_impute()} and \code{plot\_errors()}, are used to visualize results and error summaries for the imputation methods.  The package also allows users to include additional imputation methods and error functions as needed.

\subsection{The \code{impute\_errors()} function:}

The \code{impute\_errors()} function includes thirteen arguments as discussed below. This function evaluates the precision of different imputation methods based on changes in the amount and type of missing observations from the complete dataset. The default imputation functions included in \code{impute\_errors()} are \code{na.approx()} (\CRANpkg{zoo}), \code{na.interp()} (\CRANpkg{forecast}), \code{na.interpolation()} (\CRANpkg{imputeTS}), \code{na.locf()} (\CRANpkg{zoo}), and \code{na.mean()} (\CRANpkg{imputeTS}).   None of the arguments are required since all of them include default or \code{NULL} values. The syntax is shown below.

\begin{example}
impute_errors(dataIn = NULL, smps = "mcar", methods = c("na.approx",
  "na.interp", "na.interpolation", "na.locf", "na.mean"), methodPath = NULL,
  errorParameter = "rmse", errorPath = NULL, blck = 50, blckper = TRUE,
  missPercentFrom = 10, missPercentTo = 90, interval = 10,
  repetition = 10, addl_arg = NULL)
\end{example}

\subsubsection{\code{dataIn}:}
A \code{ts} (\CRANpkg{stats}) object that will be evaluated. The input object is a complete dataset with no missing values.  Missing observations are generated randomly for performance evaluation and comparison of imputation methods.  The default dataset if \code{dataIn = NULL} is \code{nottem}, a time series object of average air temperatures recorded at Nottingham Castle from 1920 to 1930.  This dataset is included with the base \CRANpkg{datasets} package.

\subsubsection{\code{smps}:}
The desired type of sampling method for removing values from the complete time series provided by \code{dataIn}. Options are \code{smps = 'mcar'} for missing completely at random (default) and \code{smps = 'mar'} for missing at random. Both methods provide completely different approaches to generating missing data in time series, as described below.

\subsubsection{\code{methods}:}
Methods that are used to impute the missing values generated by \code{smps}. All five default methods are used unless the argument is changed by the user.  For example, \code{methods =  'na.approx'} will use only \code{na.approx()} with \code{impute\_errors()}.  Methods not included with the default options can be added by including the name of the function in \code{methods} and providing the path to the script in \code{methodPath}. Additional arguments passed to each method can be included in \code{addl\_arg} described below.

\subsubsection{\code{methodPath}:}
A character string for the path of the user-supplied script that includes one to many methods passed to \code{methods}.  The path can be absolute or relative within the current working directory for the R session.  The function sources the file indicated by \code{methodPath} to add the user-supplied function to the global environment.

\subsubsection{\code{errorParameter}:}
The error metric used to compare the true, observed values from \code{dataIn} with the imputed values. Commonly used error metrics are Root Mean Square Error (\textit{RMSE}), Mean Absolute Percent Error (\textit{MAPE}) and Mean Absolute Errors (\textit{MAE}). These measures are included with \pkg{imputeTestbench} and can be used to evaluate the imputed observations by specifying \code{errorParameter = 'rmse'} (default), \code{'mape'}, or \code{'mae'}.  Additional error measures can be provided as user-supplied functions where the first argument is observed values (numeric) and the second is the imputed values (numeric). The user-supplied function must return a single numeric value as the error measure. Examples below demonstrate the addition of a user-supplied function.

\subsubsection{\code{errorPath}:}
A character string for the path of the user-supplied script that includes one to many error methods passed to \code{errorParameter}.

\subsubsection{\code{blck}:}
The block size for missing data if the sampling method is at random, \code{smps = 'mar'}. The block size can be specified as a percentage of the total amount of missing observations in \code{interval} or as a number of time steps in the input dataset.

\subsubsection{\code{blckper}:}
A logical value indicating if the numeric value passed to \code{blck} is a percentage (\code{blckper = TRUE}) or a count of time steps (\code{blckper = FALSE}).  This argument only applies if \code{smps = 'mar'}.

\subsubsection{\code{missPercentFrom}, \code{missPercentTo}:}
The minimum and maximum percentages of missing values, respectively, that are introduced in \code{dataIn}. Appropriate values for these arguments are $10$ to $90$, indicating a range from few missing observations to almost completely absent observations.

\subsubsection{\code{interval}:}
The interval of missing data from \code{missPercentFrom} to \code{missPercentTo}. The default value is $10$\% such that missing percentages in \code{dataIn} are evaluated from $10$\% to $90$\% at an interval of $10$\%, i.e., $10$\%, $20$\%, $30$\%, ..., $90$\%.  Combined, these arguments are identifical to \code{seq(from = 10, to = 90, by = 10)}.

\subsubsection{\code{repetition}:}
The number of repetitions at each \code{interval}.  Missing values are placed randomly in the original data such that multiple repetitions must be evaluated for a robust comparison of the imputation methods.

Considering the default values, the \code{impute\_errors()} function returns an \code{errprof} object as the \textit{error profile} for the imputation methods:

\begin{example}
library(imputeTestbench)
set.seed(123)
a <- impute_errors()
a
## $Parameter
## [1] "rmse"
## 
## $MissingPercent
## [1] 10 20 30 40 50 60 70 80 90
## 
## $na.approx
## [1] 0.84 1.33 1.95 3.01 3.80 4.89 6.61 8.39 9.94
## 
## $na.interp
## [1] 0.78 1.11 1.44 1.65 1.90 2.06 2.34 2.57 2.96
## 
## $na.interpolation
## [1]  0.84  1.35  2.00  3.02  3.98  5.04  6.76  8.52 10.15
## 
## $na.locf
## [1]  1.7  2.7  3.8  5.2  6.3  7.8  9.3 10.5 11.4
## 
## $na.mean
## [1] 2.6 3.8 4.7 5.4 6.1 6.6 7.2 7.7 8.2
\end{example}

The \code{errprof} object is a list with seven elements.  The first element, \code{Parameter}, is a character string of the error metric used for comparing imputation methods. The second element, \code{MissingPercent}, is a numeric vector of the missing percentages that were evaluated in the input dataset.  The remaining five elements show the average error for each imputation method at each interval of missing data in \code{MissingPercent}.  The averages at each interval are based on the repetitions specified in the initial call to \code{impute\_errors()} where the default is \code{Repetition = 10}.  Although the print method for the \code{errprof} object returns a list, the object stores the unique error estimates for every imputation method, repetition, and missing data interval.  These values are used to estimate the averages in the printed output and to plot the distribution of errors with \code{plot\_errors()} shown below.  All error values can be accessed as follows.

\begin{example}
attr(a, 'errall')
\end{example}

\subsection{Viewing results from \code{impute\_errors()}}

The \code{plot\_errors()} function can be used to plot summaries of the error metrics for each method used in \code{impute\_errors()}. This function uses the \code{errorprof} object as input and returns a graph of error values to compare the different imputation methods across the interval of missing data.  Three plot types are provided by \code{plot\_errors()} and are specified with the \code{plotType} argument.  The default value is \code{plotType = 'boxplot'} that graphs the distribution of error values for each method and missing data interval using boxplot summaries (i.e., 25th, 50th, and 75th percentile shown by the box, whiskers as 1.5 times interquartile range, and outliers beyond).  The boxplots are created using all error values stored in the \code{'errall'} attribute of the \code{errprof} object.

\begin{example}
plot_errors(a)
\end{example}

\begin{figure}[!h]

{\centering \includegraphics[width=\textwidth]{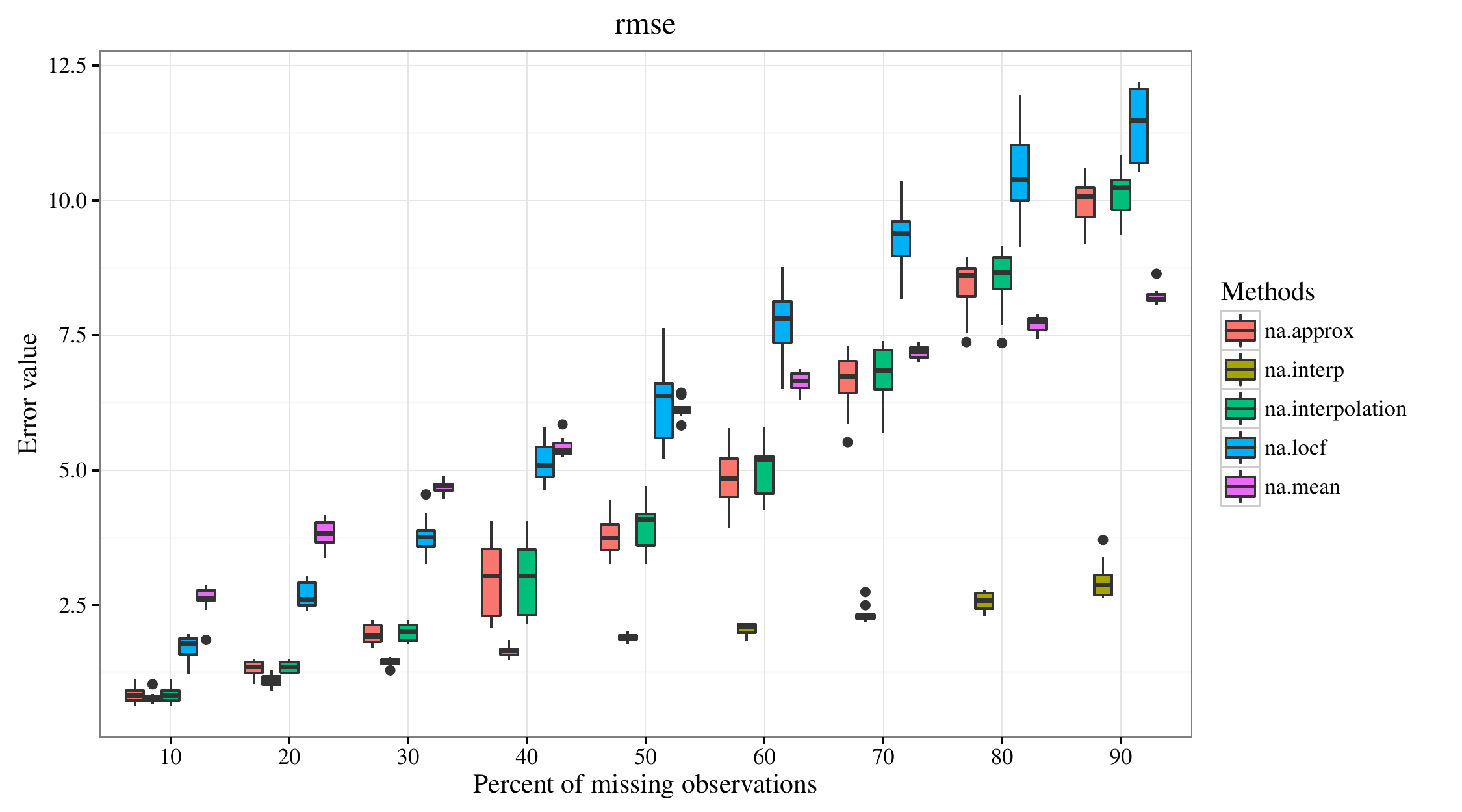} 

}

\caption[Distribution of error values for each imputation method and interval of missing observations]{Distribution of error values for each imputation method and interval of missing observations.}\label{fig:unnamed-chunk-4}
\end{figure}

The bar and line options for \code{plotType} show the average error values for each repetition.  Similar information is shown as the \code{boxplot} option, although the range of error values for each imputation method and percent of missing observations is not shown.

\begin{example}
plot_errors(a, plotType = 'line')
\end{example}

\begin{figure}[!h]

{\centering \includegraphics[width=\textwidth]{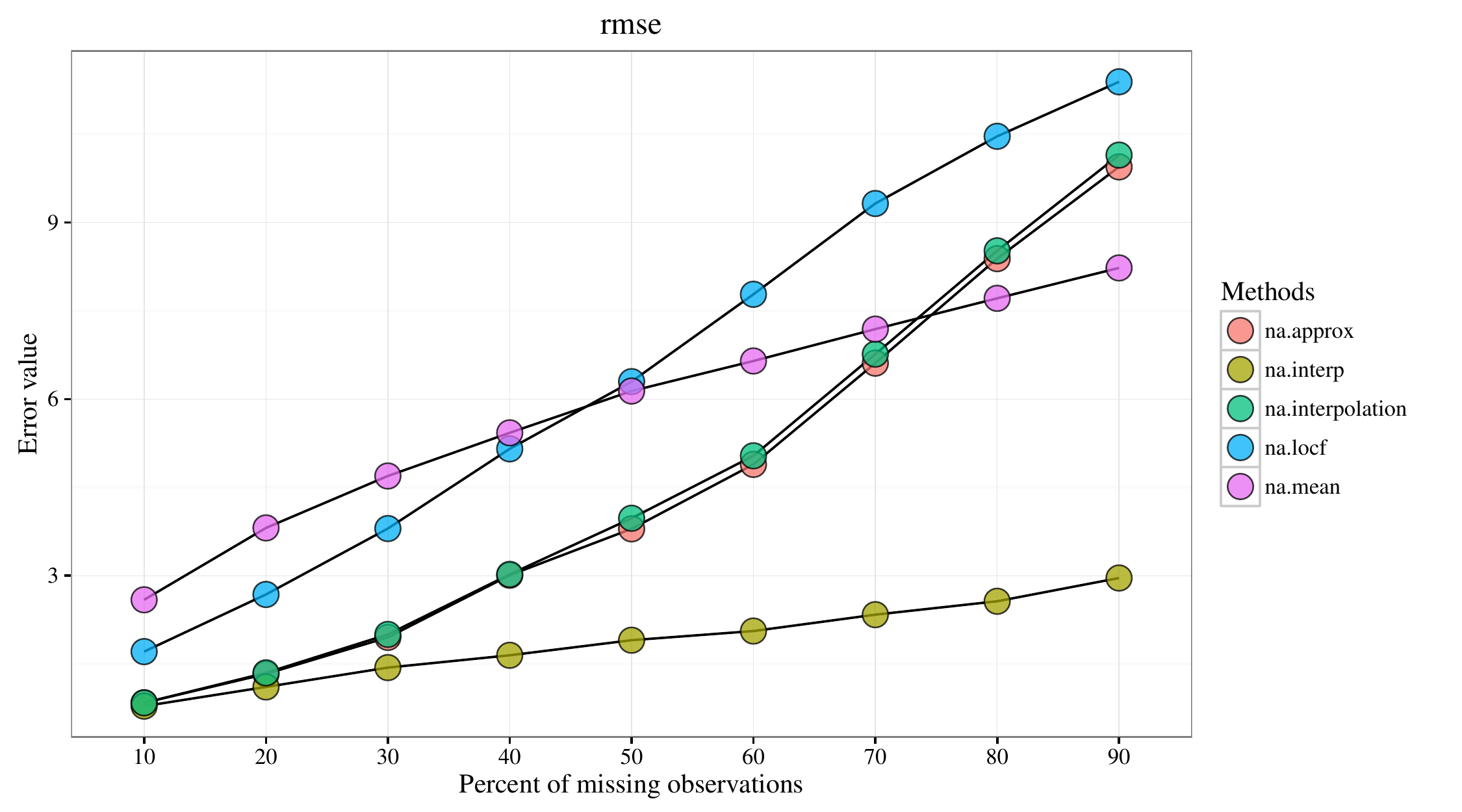} 

}

\caption[Average error values for each imputation method and interval of missing observations]{Average error values for each imputation method and interval of missing observations.  The \code{line} option is used for \code{plot\_errors()}.}\label{fig:unnamed-chunk-5}
\end{figure}

\subsection{Sampling methods for missing observations}

The \code{impute\_errors()} function uses the \code{sample\_dat()} function to remove observations for imputation from the input dataset.  The \code{sample\_dat()} function removes observations using one of two methods that are relevant for univariate time series.  Observations can be removed following a missing completely at random (MCAR) or missing at random (MAR) sampling scheme with the appropriate \code{smps} argument. The MCAR sampling scheme assumes all observations have equal probability of being selected for removal and is appropriate for univariate time series that are not serially correlated (i.e., no temporal dependence).  Conversely, the MAR sampling scheme selects observations in blocks such that the probability of selection for a single observation depends on whether an observation closer in time was also selected.  The MAR scheme is appropriate for time series with serial correlation.  For example, missing data may occur in univariate time series if monitoring equipment fail for a period of time or data are not collected on the weekends. The \code{sample\_dat} function has the following syntax:

\begin{example}
sample_dat(datin, smps = "mcar", repetition = 10, b = 50, blck = 50,
  blckper = TRUE, plot = FALSE)
\end{example}

\subsubsection{\code{datin}:}
Input numeric vector, inherited from \code{dataIn} from \code{impute\_errors}.

\subsubsection{\code{smps, repetition, blck, blckper}:} Arguments that are inherited as is from \code{impute\_errors} indicating the sampling type (\code{smps}), number of repetitions for each missing data type (\code{repetition}), block size (\code{blck}), and block type as percentage or count (\code{blckper}).

\subsubsection{\code{b}:}
Numeric indicating the total amount of missing data as a percentage to remove from the complete time series. The values passed to \code{b} within \code{impute\_errors} are those defined by \code{missPercentFrom}, \code{missPercentTo}, and \code{interval}.

\subsubsection{\code{plot}:}
Logical indicating if a plot is returned that shows one repetition of the sampling scheme defined by the arguments (see Figure 4). 

The MCAR sampling scheme is used if \code{smps = 'mcar'} in the call to \code{impute\_errors()}.  The only relevant arguments for MCAR are \code{missPercentFrom}, \code{missPercentTo}, and \code{interval} that define the amount of data to remove as a percentage of the total.  The amount of data to remove for each interval is passed to the \code{b} argument in \code{sample\_dat}.  The MAR sampling scheme requires additional arguments to control the block size for removing data in continuous chunks, in addition to the total amount of data to remove.  The block size argument, \code{blck}, can be given as a percentage or as number of observations in sequence.  The type of block size passed to \code{blck} is controlled by \code{blckper}, where \code{blckper = TRUE} indicates a percentage and \code{FALSE} indicates a count for \code{blck}.  For example, if the total sample size of the dataset is 1000, \code{b = 50}, \code{blck = 20}, and \code{blckper = TRUE} means half the dataset is removed (\code{b = 50}, 500 observations) and each block will have 100 observations (20\% of 500).  For both percentages and counts, the blocks are automatically selected until the total amount of missing data is equal to that specified by \code{b}.  Final blocks may be truncated to make the total amount of missing observations equal to \code{b}.  The starting location of each block is selected at random and overlapping blocks are not uniquely counted for the required sample size given by \code{b}.

The \code{sample\_dat()} function includes an optional \code{plot} argument.  Although the function is primarily used within \code{impute\_errors} to generate missing data, it can be used independently to visualize different sampling schemes.  Figure 4 shows some examples of sampling completely at random (MCAR) and at random (MAR).

\begin{figure}[h!]

{\centering \includegraphics[width=\textwidth]{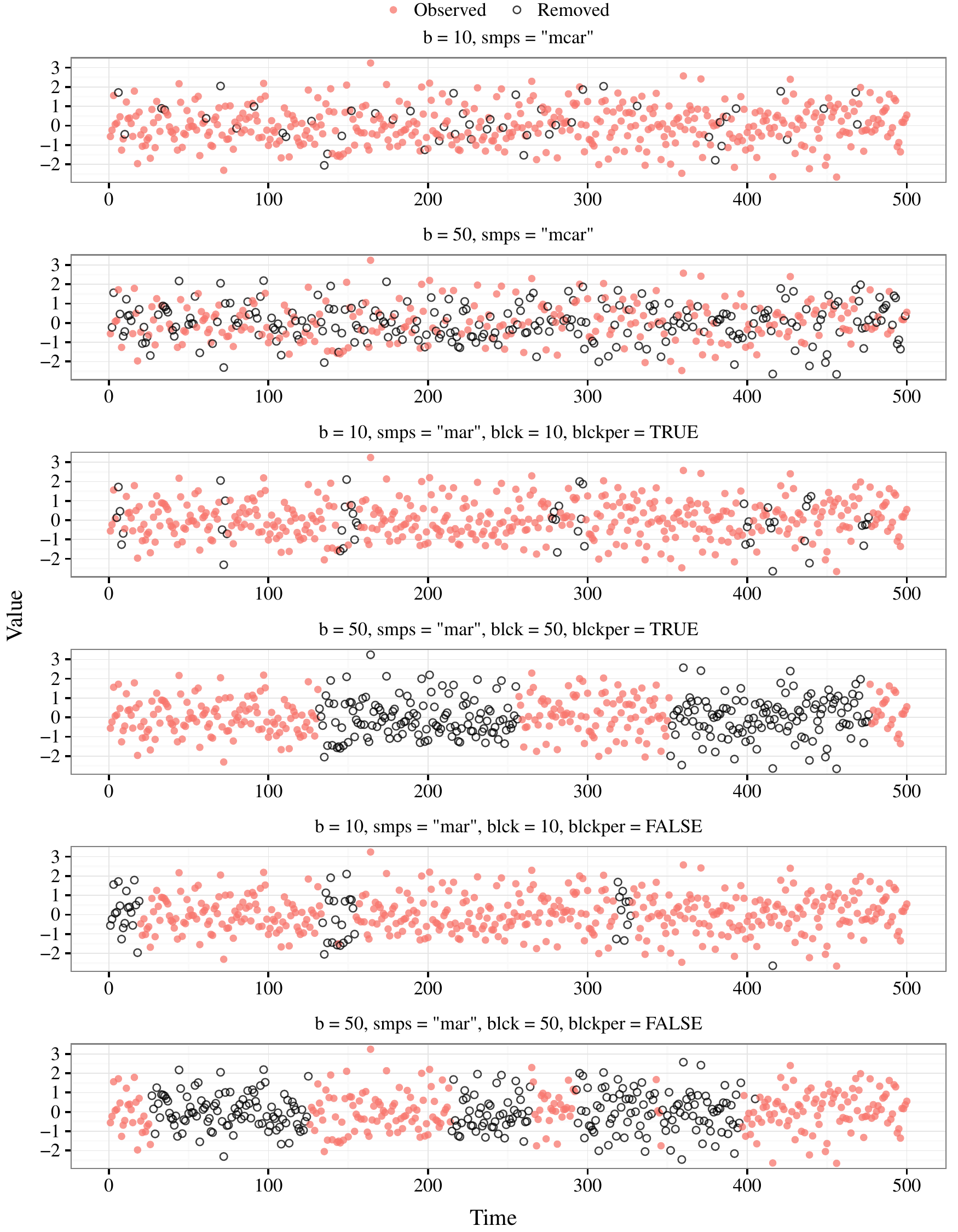}}

\caption[Examples of sampling schemes for missing data provided by \code{sample\_dat()}]{Examples of sampling schemes for missing data provided by \code{sample\_dat()}.  Values to be removed and imputed are shown as open circles and the data to be kept are in red. From top to bottom, sampling is completely at random (MCAR) with 10\% missing, MCAR with 50\% missing, sampling at random (MAR) with 10\% missing and block size 10\% of total missing, MAR with 50\% missing and block size 50\% of total missing, MAR with 10\% missing and block size of ten observations, and MAR with 50\% missing and block size of fifty observations.}\label{fig:unnamed-chunk-8}
\end{figure}

\clearpage

\subsection{The \code{plot\_impute()} function}

An additional plotting function available in \pkg{imputeTestbench} is \code{plot\_impute()}.  This function returns a plot of the imputed values for each imputation method in \code{impute\_errors()} for one repetion of sampling with \code{sample\_dat()}.  The plot shows the results as a single facet for each method with the points colored as not filled or filled (i.e., original data not removed and filled data that were removed).  An optional argument, \code{showmiss}, can be used to show the original values as open circles that were removed from the data.  It should be noted that the plot from \code{plot\_errors()} is a more accurate representation of the abilities of each method.  The \code{plot\_impute()} function shows results for only one simulation and missing data type (e.g., \code{smps = 'mcar'} and \code{b = 50}). This function is useful as a simple visualization of the sampling scheme for the missing values and the relative abilities of each method for imputation.
\begin{example}
plot_impute(showmiss = T)
\end{example}

\begin{figure}[h!]

{\centering \includegraphics[width=\textwidth]{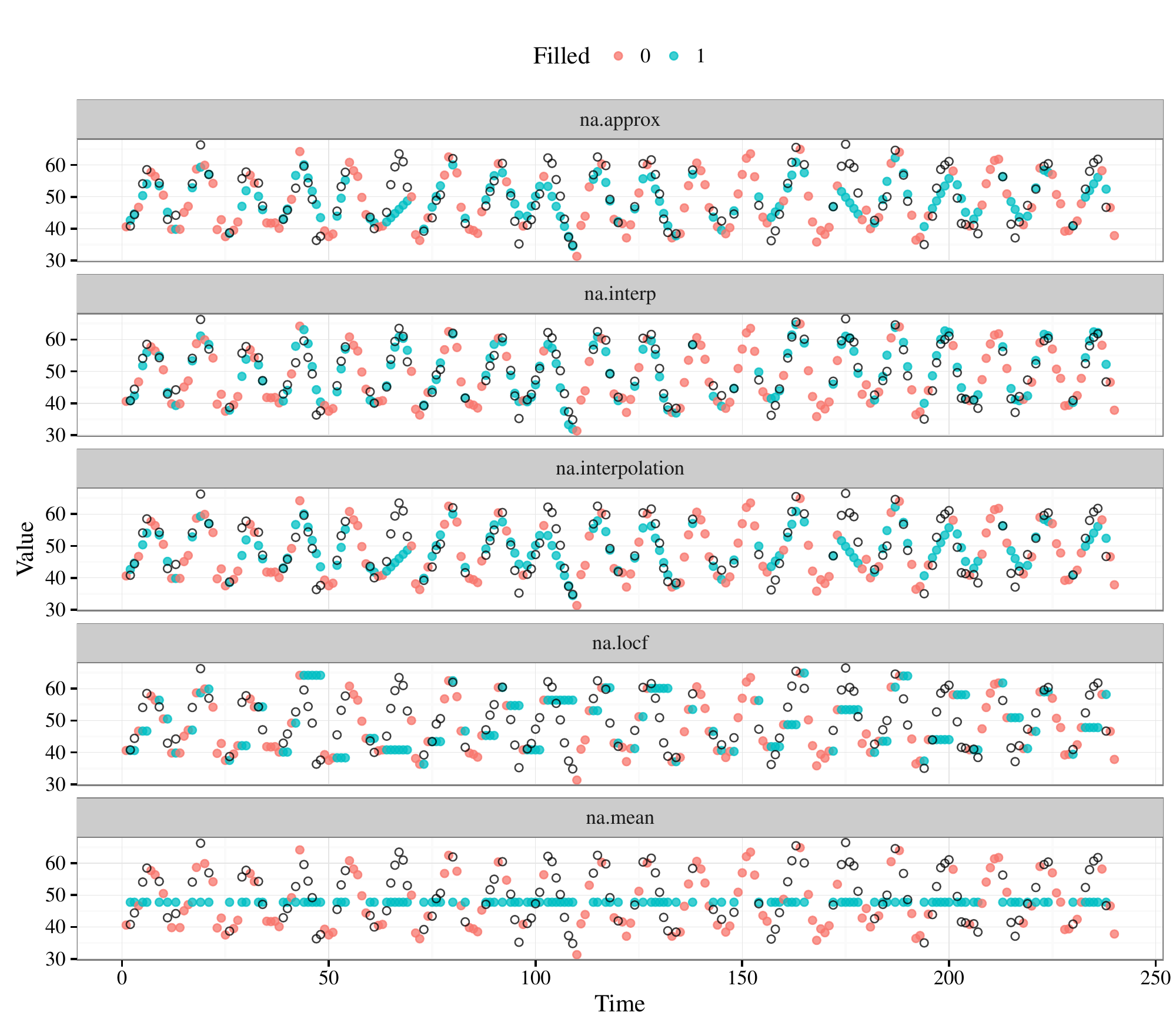} 

}

\caption[Output from the \code{plot\_impute()} function that shows the data that were not removed (red), removed (open circles), and imputed (blue)]{Output from the \code{plot\_impute()} function that shows the data that were not removed (red), removed (open circles), and imputed (blue).}\label{fig:unnamed-chunk-9}
\end{figure}

\clearpage

\subsection{Adding error metrics and imputation methods}

The \code{error\_functions()} function is a collection of error metrics that are used to evaluate differences between the original and imputed data. This function is used internally within \code{impute\_errors()} to compare results from the imputation methods.  As described above, the available error metrics are \textit{RMSE}, \textit{MAE}, and \textit{MAPE}. However, the proposed testbench is not limited to these metrics and alternative functions can be provided by the user.  An alternative error metric can be used by specifying the path to the function using the \code{errorPath} argument in \code{impute\_errors()}.  The \code{errorParameter} argument must also be changed to the name of the function in \code{errorPath}.  The user-supplied function should accept two arguments as input, the first being the observed time series and the second being the imputed data for comparison.  The function must return a single numeric value that is the result of comparing the two input vectors.  

Similarly, imputation methods can be added to \code{impute\_errors()} by providing a path for an R script to the \code{methodPath} argument.  The added imputation method must also be added as a character string to the \code{methods} argument.  User-supplied imputation functions should have one required argument for the input \code{ts} object and should return an imputed vector of observations of the same length as the input object.  Examples of adding error metrics and additional imputation methods are provided in the next section. 

Additional arguments for user-supplied imputation functions can also be included.  These arguments can be passed to \code{impute\_errors} using the \code{addl\_arg} argument, as can any arguments for the default imputation methods.  The additional arguments are passed as a \code{list} of lists to the \code{addl\_arg} argument, where the list contains one to many elements that are named by the methods in \code{methods}. The elements of the list are lists with arguments that are specific to each imputation method.  For example, the default function \code{na.mean} has an additional \code{option} argument that specifies the algorithm to use for missing values, where possible values are \code{"mean"}, \code{"median"}, and \code{"mode"}.  This argument can be changed from the default \code{option = "mean"} with \code{addl\_arg} in \code{impute\_errors}, as shown below. Arguments to user-supplied imputation functions can be changed similarly.
\begin{example}
# changing the option argument for na.mean
impute_errors(addl_arg = list(na.mean = list(option = 'mode')))
\end{example}

\section{Demonstration of \pkg{imputeTestbench} with examples}

This example demonstrates how \pkg{imputTestbench} can be used to compare different imputation methods. The testbench is always initiated with the \code{impute\_errors()} function, which uses time series data and additional arguments as discussed in the previous section. The example uses the \code{austres} dataset available in the \CRANpkg{datasets} package. The \code{impute\_errors()} function with the \textit{RMSE} error metric and five default imputation methods returns the error profile below:

\begin{example}
aus <- datasets::austres
ex <- impute_errors(dataIn = aus)
ex
## $Parameter
## [1] "rmse"
## 
## $MissingPercent
## [1] 10 20 30 40 50 60 70 80 90
## 
## $na.approx
## [1]  1.9  3.1  3.6  5.1  6.0  8.5 12.2 19.7 31.0
## 
## $na.interp
## [1]   2.4   3.8   9.8  14.8  20.1  24.1  29.1  55.0 191.7
## 
## $na.interpolation
## [1]   2.5   3.8   9.3  12.4  16.3  20.0  23.4  44.7 177.1
## 
## $na.locf
## [1]  19  32  47  68  89 123 168 286 574
## 
## $na.mean
## [1]  430  611  759  867  936 1064 1123 1257 1380
plot_errors(ex, plotType = 'line')
\end{example}

\begin{figure}[!h]

{\centering \includegraphics[width=\textwidth]{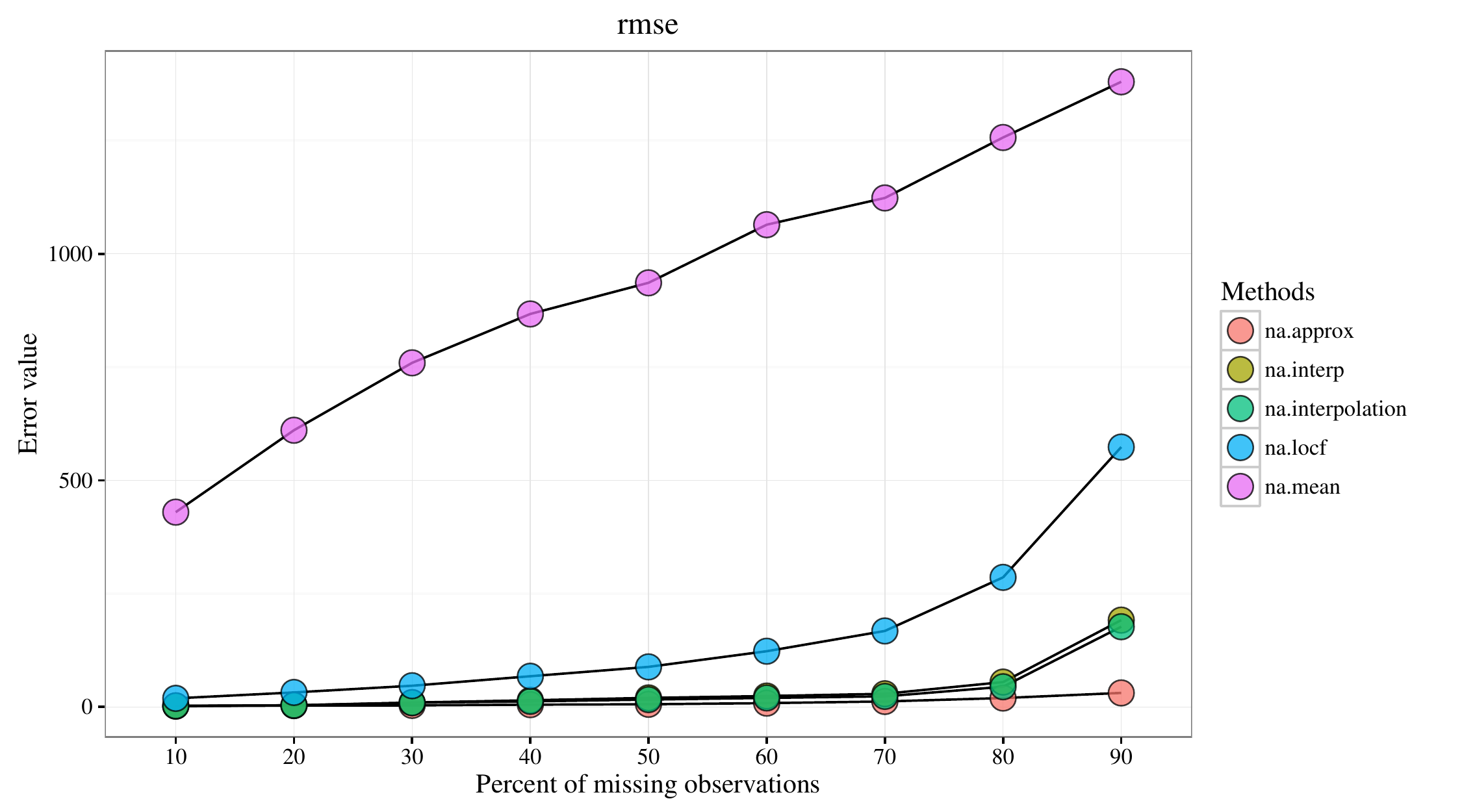} 

}

\caption[RMSE comparison of imputations using the \code{austres} dataset]{RMSE comparison of imputations using the \code{austres} dataset.}\label{fig:unnamed-chunk-11}
\end{figure}

The \code{austres} dataset is a \code{ts} object of Australian population in thousands, measured quarterly from 1971 to 1994 \citep{brockwell96}.  The \code{plot\_errors()} function shows that all imputation methods had larger error values with additional missing observations, as expected, and that the \code{na.mean} imputation method had the largest error values.  Differences between the error values can be understood by viewing a sample of the imputed data with \code{plot\_impute()}.  The example belows shows an example of imputed values using the \code{na.approx}, \code{na.locf}, and \code{na.mean} functions at 10\% and 90\% missing observations using MCAR sampling.  
\begin{figure}[!h]

{\centering \includegraphics[width=\textwidth]{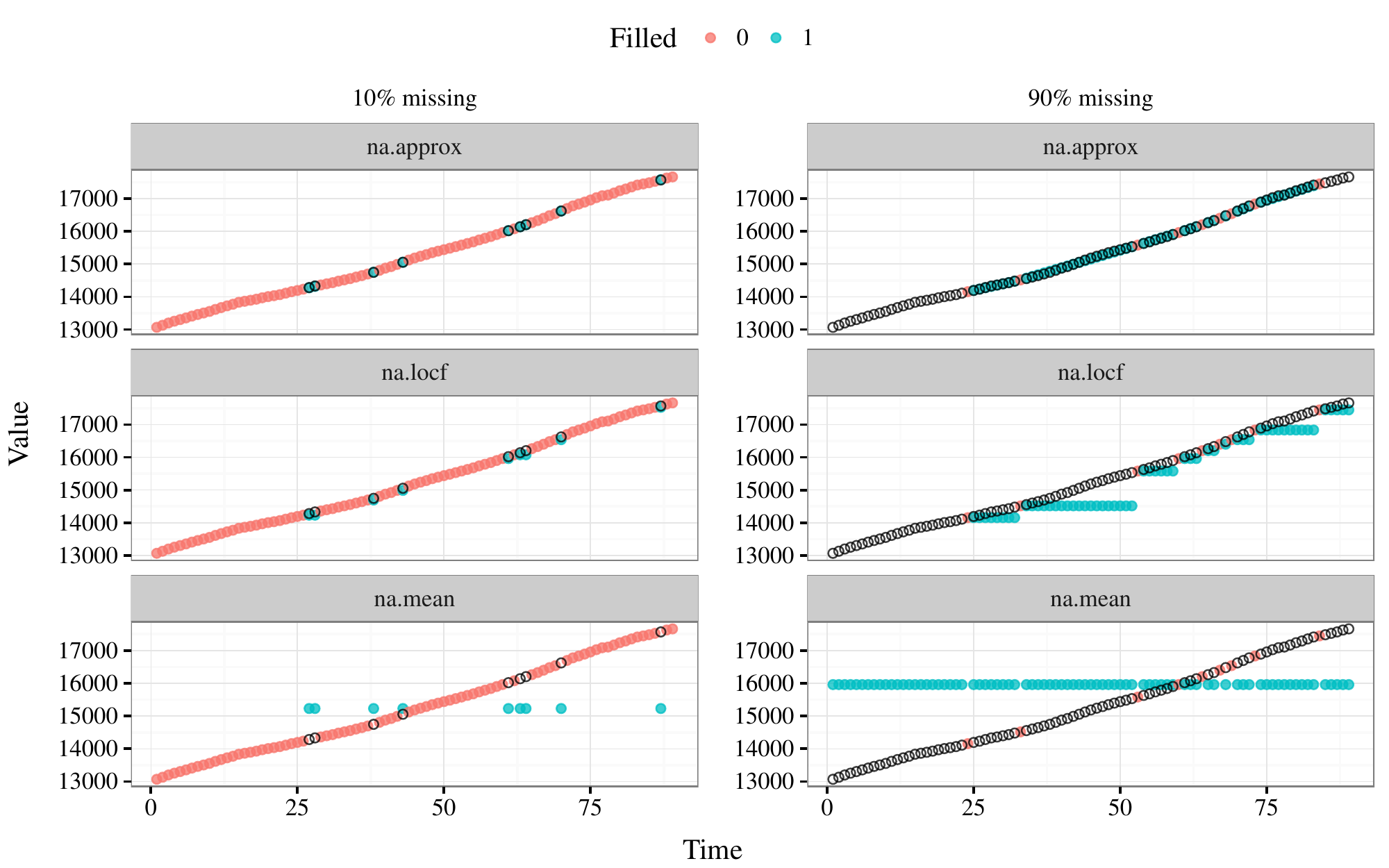} 

}

\caption{An example from \code{plot\_impute()} of imputed values for the \code{austres} dataset using \code{na.approx}, \code{na.locf}, and \code{na.mean}.  The left and right columns shows 10\% and 90\% missing data with MCAR sampling.}\label{fig:unnamed-chunk-12}
\end{figure}

Reasons for differences in error values between the methods are apparent from \code{plot\_impute}.  The \code{austres} data is a serially correlated time series that increases linearly throughout the time period.  The \code{na.mean} function performs poorly because it does not capture the linear increase through time.  Conversely, \code{na.locf} and \code{na.approx} perform equally well for small percentages of missing data but error values diverge for larger percentages.  These trends are shown in Figure 6 and verified in Figure 7.  As such, differences in error values between methods relate to the characteristics of the dataset and the interpolation method used by each function.

As described above, imputation methods supplied by the user can be added to \code{impute\_errors()}. The example below demonstrates the addition of a random number imputation method to the error profile. An R script file must be created for adding and saving the function. Additional functions can be added to the script as needed.  User-supplied functions for imputation should use time series data with missing values as input and return the time series data with the imputed values as shown below.

\begin{example}
# A sample function to randomly impute the missing data
library(imputeTS)
sss <- function(In){
  out <- na.random(In)
  out <- as.numeric(out)
  return(out)
}
\end{example}

The path where the R script is saved is used as an input string to the \code{methodPath} argument.  The name of the new function is added to the \code{methods} argument, including any of the default methods used by \code{impute\_errors}.  Results are shown below and in Figure 8.

\begin{example}
ex <- impute_errors(dataIn = aus, methodPath = 'SupportiveCodes/sss.R',
                        methods = c('na.mean', 'na.locf', 'na.approx', 'sss'))
ex
## $Parameter
## [1] "rmse"
## 
## $MissingPercent
## [1] 10 20 30 40 50 60 70 80 90
## 
## $na.mean
## [1]  401  594  763  894  936 1034 1150 1255 1353
## 
## $na.locf
## [1]  19  33  49  66  89 111 184 263 477
## 
## $na.approx
## [1]  1.6  2.7  3.9  4.7  6.0  7.4 10.5 16.1 26.6
## 
## $sss
## [1]  582  795  978 1218 1221 1549 1497 1414 1674
plot_errors(ex, plotType = 'line')
\end{example}

\begin{figure}[!h]

{\centering \includegraphics[width=\textwidth]{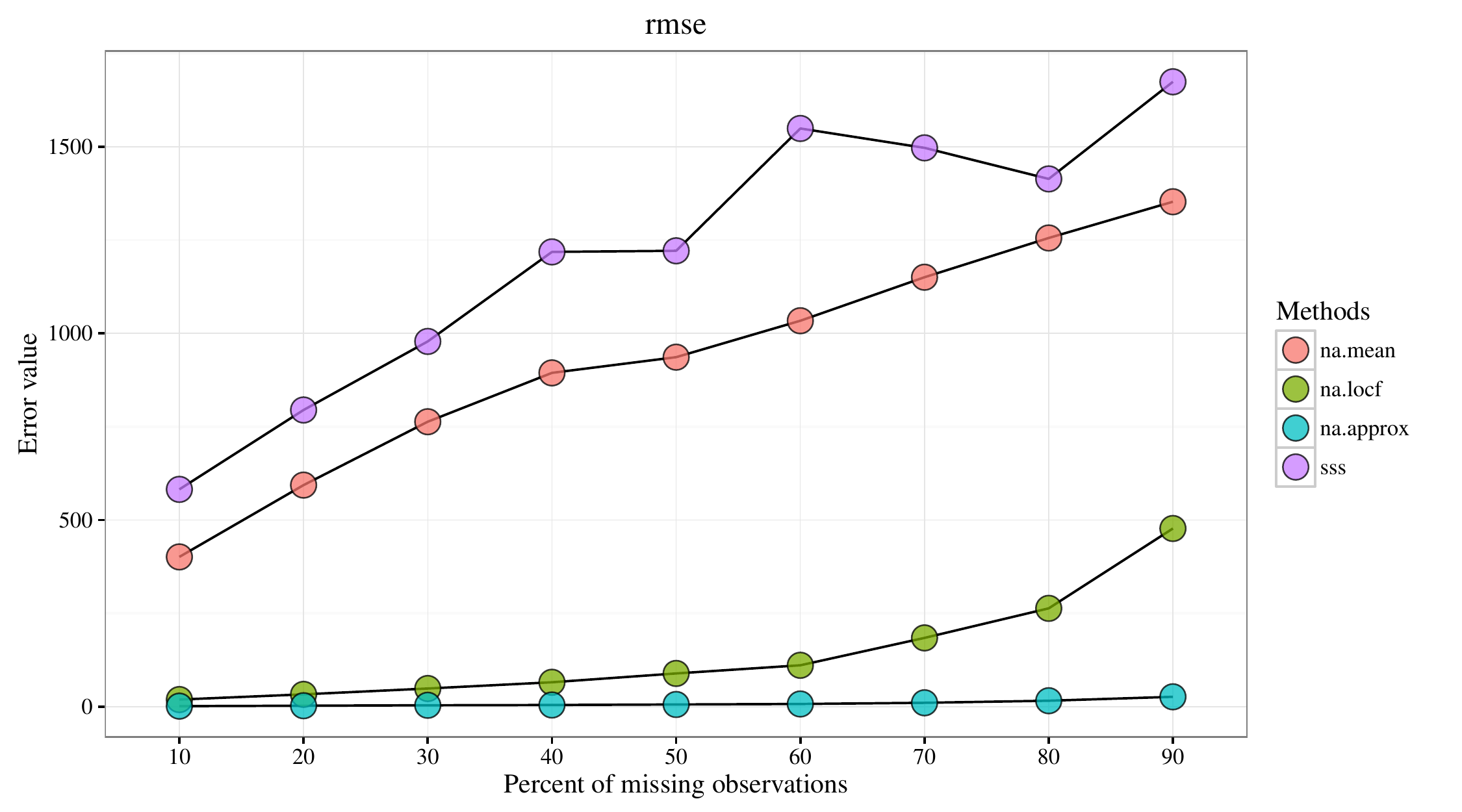} 

}

\caption[Adding a new imputation method to fill missing data in the \code{austres} dataset]{Adding a new imputation method to fill missing data in the \code{austres} dataset.}\label{fig:unnamed-chunk-14}
\end{figure}

An error metric can be added similarly. The following example shows use of the percent change in variance \citep[\textit{PCV},][]{tak7444178} as an alternative error metric:

\begin{equation}
{PCV} = \frac{var(\bar{V}) - var(V)}{var(V)}\
\end{equation}

Error is estimated as the difference between the variance of the imputed data, $var(\bar{V})$, and variance of the missing data, $var(V)$, divided by the variance of the missing data.  The user-supplied error function must include two arguments as input, the first being a vector of observed values and the second being a vector of imputed missing values equal in length to the first.  The function must also return a single value as a summary of the errors or differences.
\begin{example}
# the pcv error function 
pcv <- function(dataIn, imputed)
{
  d <- (var(imputed) - var(dataIn)) * 100/ var(imputed)
  d <- as.numeric(d)
  return(d)
}
\end{example}

As before, the new error function should be saved as an R script. The file path is added to the \code{errorPath} argument and the error function name is added to the \code{errorParameter} argument for the \code{impute\_errors} function. Results are shown below and in Figure 9.

\begin{example}
ex <- impute_errors(dataIn = aus, errorPath = 'SupportiveCodes/pcv.R', 
  errorParameter = 'pcv')
ex
## $Parameter
## [1] "pcv"
## 
## $MissingPercent
## [1] 10 20 30 40 50 60 70 80 90
## 
## $na.approx
## [1]  -0.49  -1.56  -0.75  -2.38  -2.91 -10.86  -9.93 -23.28 -66.28
## 
## $na.interp
## [1]  -0.0048  -0.1405  -0.0333  -0.3056  -0.3367  -1.8588  -1.5596
## [8]  -5.0646 -22.2181
## 
## $na.interpolation
## [1]  -0.0092  -0.1337  -0.0252  -0.2842  -0.3964  -1.6820  -1.4542
## [8]  -4.9024 -23.7908
## 
## $na.locf
## [1]   0.083  -1.143  -0.198  -1.451  -3.230  -4.726  -6.559  -9.041
## [9] -42.841
## 
## $na.mean
## [1]   -11   -29   -43   -69  -110  -166  -258  -452 -1128
plot_errors(ex, plotType = 'line')
\end{example}

\begin{figure}[!h]

{\centering \includegraphics[width=\textwidth]{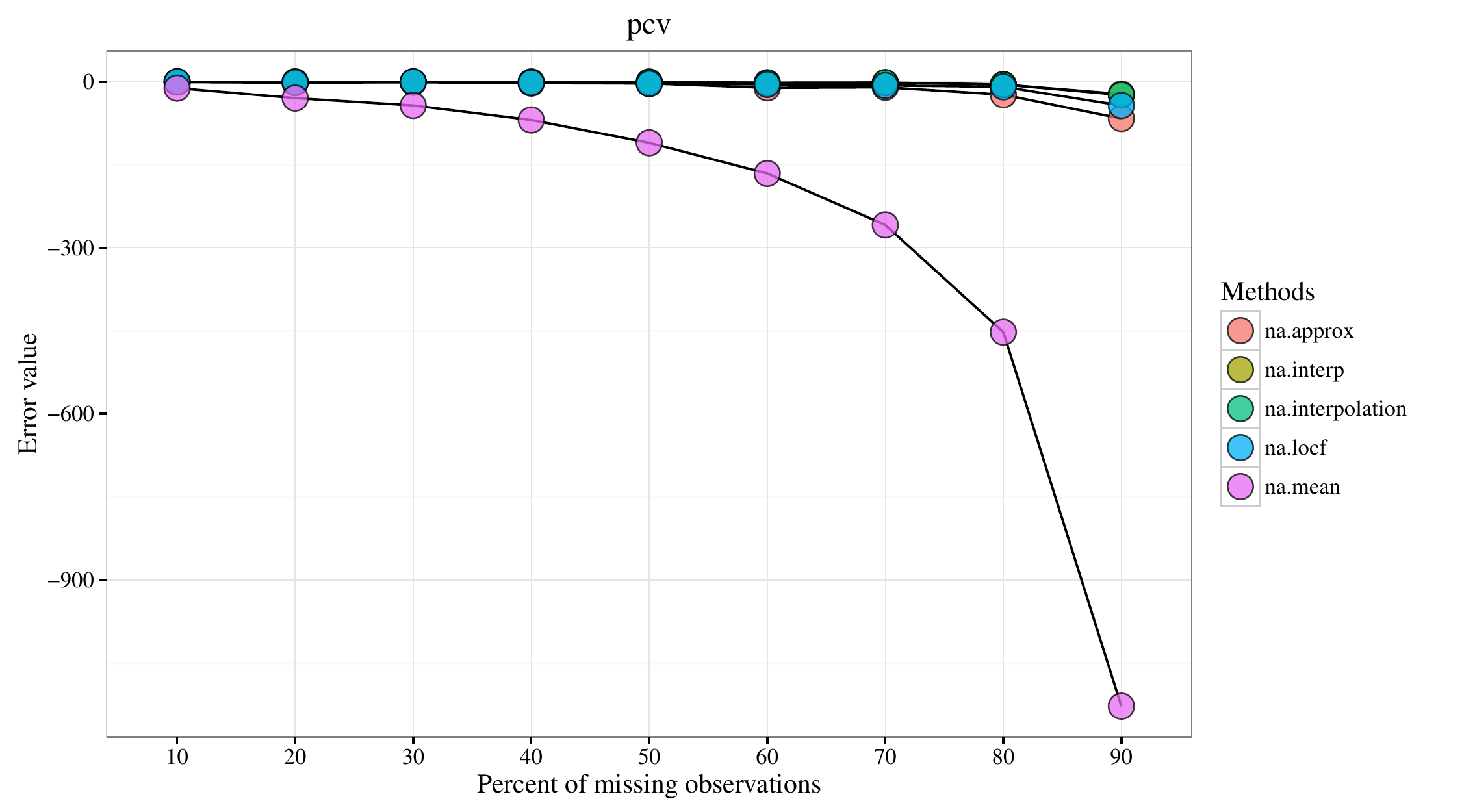} 

}

\caption[Adding a new error parameter to compare imputation methods for missing data in the \code{austres} dataset]{Adding a new error parameter to compare imputation methods for missing data in the \code{austres} dataset.}\label{fig:unnamed-chunk-16}
\end{figure}

\section{Summary}

This paper described the \pkg{imputeTestbench} \citep{imputeTB} R package which works as a testbench to compare imputation methods for missing data. The usability of this package was demonstrated by the examples above. By default, the testbench compares existing imputation methods (\code{na.approx()}, \code{na.interp()}, \code{na.interpolation()}, \code{na.locf()}, and \code{na.mean()}) using \textit{RMSE}, \textit{MAE} or \textit{MAPE} error metrics. Along with the default methods, the package allows users to include additional imputation methods for comparison. As such, this package can support imputation methods compiled with C, Fortran, C++, Java, Python or Matlab languages with the help of R packages like \CRANpkg{Rcpp} \citep{rcpp}, \CRANpkg{rJava} \citep{rjava}, \CRANpkg{rPython}, and \CRANpkg{matlabr} \citep{matlabr}. Imputation methods can also be evaluated with alternative error metrics other than those provided with the package.  As such, the simple architecture of \pkg{imputeTestbench} to add or remove multiple methods and error metrics makes it a robust and useful tool to evaluate existing and proposed imputation techniques. The results discussed in this paper are performed using R 3.3.1. The \pkg{imputeTestbench} package is available on CRAN (\href{https://cran.r-project.org/web/packages/imputeTestbench/index.html]}{https://cran.r-project.org/web/packages/imputeTestbench/index.html}). 

\bibliography{bokde-kulat}

\begin{thebibliography}{25}
\providecommand{\natexlab}[1]{#1}
\providecommand{\url}[1]{\texttt{#1}}
\expandafter\ifx\csname urlstyle\endcsname\relax
  \providecommand{\doi}[1]{doi: #1}\else
  \providecommand{\doi}{doi: \begingroup \urlstyle{rm}\Url}\fi

\bibitem[Bokde and Beck(2016)]{imputeTB}
N.~Bokde and M.~W. Beck.
\newblock \emph{imputeTestbench: Test Bench for Missing Data Imputing
  Models/Methods Comparison}, 2016.
\newblock URL \url{https://cran.r-project.org/package=imputeTestbench}.
\newblock R package version 3.0.0.

\bibitem[Brockwell and Davis(1996)]{brockwell96}
P.~J. Brockwell and R.~A. Davis.
\newblock \emph{Introduction to Time Series and Forecasting}.
\newblock Springer-Verlag New York, 1996.
\newblock ISBN 978-1-4757-2526-1.

\bibitem[Buuren and Groothuis-Oudshoorn(2011)]{buuren2011mice}
S.~Buuren and K.~Groothuis-Oudshoorn.
\newblock mice: Multivariate imputation by chained equations in r.
\newblock \emph{Journal of statistical software}, 45\penalty0 (3), 2011.

\bibitem[Eddelbuettel and Fran\c{c}ois(2011)]{rcpp}
D.~Eddelbuettel and R.~Fran\c{c}ois.
\newblock {Rcpp}: Seamless {R} and {C++} integration.
\newblock \emph{Journal of Statistical Software}, 40\penalty0 (8):\penalty0
  1--18, 2011.
\newblock URL \url{http://www.jstatsoft.org/v40/i08/}.

\bibitem[Honaker et~al.()Honaker, King, Blackwell, et~al.]{honaker2011amelia}
J.~Honaker, G.~King, M.~Blackwell, et~al.
\newblock Amelia ii: A program for missing data.

\bibitem[Hyndman(2016)]{forecast}
R.~J. Hyndman.
\newblock \emph{forecast: Forecasting functions for time series and linear
  models}, 2016.
\newblock URL \url{http://github.com/robjhyndman/forecast}.
\newblock R package version 7.2.

\bibitem[J{\"o}rnsten et~al.(2007)J{\"o}rnsten, Ouyang, and
  Wang]{jornsten2007meta}
R.~J{\"o}rnsten, M.~Ouyang, and H.-Y. Wang.
\newblock A meta-data based method for dna microarray imputation.
\newblock \emph{BMC bioinformatics}, 8\penalty0 (1):\penalty0 109, 2007.

\bibitem[Li et~al.(2004)Li, Deogun, Spaulding, and Shuart]{li2004towards}
D.~Li, J.~Deogun, W.~Spaulding, and B.~Shuart.
\newblock Towards missing data imputation: a study of fuzzy k-means clustering
  method.
\newblock In \emph{Rough sets and current trends in computing}, pages 573--579.
  Springer, 2004.

\bibitem[Li et~al.(2015)Li, Zhao, Shao, Li, and Wang]{li2015hybrid}
H.~Li, C.~Zhao, F.~Shao, G.-Z. Li, and X.~Wang.
\newblock A hybrid imputation approach for microarray missing value estimation.
\newblock \emph{BMC genomics}, 16\penalty0 (Suppl 9):\penalty0 S1, 2015.

\bibitem[Moritz(2015)]{imputeTS}
S.~Moritz.
\newblock \emph{imputeTS: Time Series Missing Value Imputation}, 2015.
\newblock URL \url{https://CRAN.R-project.org/package=imputeTS}.
\newblock R package version 0.4.

\bibitem[Muschelli(2015)]{matlabr}
J.~Muschelli.
\newblock \emph{matlabr: An Interface for MATLAB using System Calls}, 2015.
\newblock URL \url{https://CRAN.R-project.org/package=matlabr}.
\newblock R package version 1.1.

\bibitem[Nguyen et~al.(2013)Nguyen, Carlin, and Lee]{nguyen2013diagnosing}
C.~D. Nguyen, J.~B. Carlin, and K.~J. Lee.
\newblock Diagnosing problems with imputation models using the
  kolmogorov-smirnov test: a simulation study.
\newblock \emph{BMC medical research methodology}, 13\penalty0 (1):\penalty0 1,
  2013.

\bibitem[Oh et~al.(2011)Oh, Kang, Brock, and Tseng]{oh2011biological}
S.~Oh, D.~D. Kang, G.~N. Brock, and G.~C. Tseng.
\newblock Biological impact of missing-value imputation on downstream analyses
  of gene expression profiles.
\newblock \emph{Bioinformatics}, 27\penalty0 (1):\penalty0 78--86, 2011.

\bibitem[Ran et~al.(2015)Ran, Tan, Feng, Liu, and Wang]{ran2015traffic}
B.~Ran, H.~Tan, J.~Feng, Y.~Liu, and W.~Wang.
\newblock Traffic speed data imputation method based on tensor completion.
\newblock \emph{Computational intelligence and neuroscience}, 2015:\penalty0
  22, 2015.

\bibitem[Sim et~al.(2015)Sim, Lee, and Kwon]{sim2015missing}
J.~Sim, J.~S. Lee, and O.~Kwon.
\newblock Missing values and optimal selection of an imputation method and
  classification algorithm to improve the accuracy of ubiquitous computing
  applications.
\newblock \emph{Mathematical Problems in Engineering}, 2015, 2015.

\bibitem[Stekhoven and B{\"u}hlmann(2012)]{stekhoven2012missforest}
D.~J. Stekhoven and P.~B{\"u}hlmann.
\newblock Missforest—non-parametric missing value imputation for mixed-type
  data.
\newblock \emph{Bioinformatics}, 28\penalty0 (1):\penalty0 112--118, 2012.

\bibitem[Su et~al.(2011)Su, Yajima, Gelman, and Hill]{su2011multiple}
Y.-S. Su, M.~Yajima, A.~E. Gelman, and J.~Hill.
\newblock Multiple imputation with diagnostics (mi) in r: Opening windows into
  the black box.
\newblock \emph{Journal of Statistical Software}, 45\penalty0 (2):\penalty0
  1--31, 2011.

\bibitem[Tak et~al.(2016)Tak, Woo, and Yeo]{tak7444178}
S.~Tak, S.~Woo, and H.~Yeo.
\newblock Data-driven imputation method for traffic data in sectional units of
  road links.
\newblock \emph{IEEE Transactions on Intelligent Transportation Systems},
  PP\penalty0 (99):\penalty0 1--10, 2016.
\newblock ISSN 1524-9050.
\newblock \doi{10.1109/TITS.2016.2530312}.

\bibitem[Urbanek(2016)]{rjava}
S.~Urbanek.
\newblock \emph{rJava: Low-Level R to Java Interface}, 2016.
\newblock URL \url{https://CRAN.R-project.org/package=rJava}.
\newblock R package version 0.9-8.

\bibitem[Wickham(2007)]{reshape2}
H.~Wickham.
\newblock Reshaping data with the {reshape} package.
\newblock \emph{Journal of Statistical Software}, 21\penalty0 (12):\penalty0
  1--20, 2007.
\newblock URL \url{http://www.jstatsoft.org/v21/i12/}.

\bibitem[Wickham(2009)]{ggplot2}
H.~Wickham.
\newblock \emph{ggplot2: Elegant Graphics for Data Analysis}.
\newblock Springer-Verlag New York, 2009.
\newblock ISBN 978-0-387-98140-6.
\newblock URL \url{http://ggplot2.org}.

\bibitem[Wickham(2016)]{tidyr}
H.~Wickham.
\newblock \emph{tidyr: Easily Tidy Data with `spread()` and `gather()`
  Functions}, 2016.
\newblock URL \url{https://CRAN.R-project.org/package=tidyr}.
\newblock R package version 0.6.0.

\bibitem[Wickham and Francois(2016)]{dplyr}
H.~Wickham and R.~Francois.
\newblock \emph{dplyr: A Grammar of Data Manipulation}, 2016.
\newblock URL \url{https://CRAN.R-project.org/package=dplyr}.
\newblock R package version 0.5.0.

\bibitem[Zeileis and Grothendieck(2005)]{zoo}
A.~Zeileis and G.~Grothendieck.
\newblock zoo: S3 infrastructure for regular and irregular time series.
\newblock \emph{Journal of Statistical Software}, 14\penalty0 (6):\penalty0
  1--27, 2005.
\newblock \doi{10.18637/jss.v014.i06}.

\bibitem[Zhu et~al.(2011)Zhu, Zhang, Jin, Zhang, and Xu]{zhu2011missing}
X.~Zhu, S.~Zhang, Z.~Jin, Z.~Zhang, and Z.~Xu.
\newblock Missing value estimation for mixed-attribute data sets.
\newblock \emph{Knowledge and Data Engineering, IEEE Transactions on},
  23\penalty0 (1):\penalty0 110--121, 2011.

\end{thebibliography}

\address{Neeraj Bokde\\
  Visvesvaraya National Institute of Technology, Nagpur\\
  North Ambazari Road, Nagpur\\
  India\\}
\email{neeraj.bokde@students.vnit.ac.in}

\address{Kishore Kulat\\
  Visvesvaraya National Institute of Technology, Nagpur\\
  North Ambazari Road, Nagpur\\
  India\\}
\email{kdkulat@ece.vnit.ac.in}

\address{Marcus W Beck\\
  USEPA National Health and Environmental Effects Research Laboratory, Gulf Ecology Division\\
  1 Sabine Island Drive, Gulf Breeze, FL 32651\\
  USA\\}
\email{beck.marcus@epa.gov}

\address{Gualberto Asencio-Cort\'{e}s\\
  Universidad Pablo de Olavide\\
  ES-41013, Sevilla\\
  Spain\\}
\email{guaasecor@upo.es}
\end{article}

\end{document}